\documentclass[pra,showpacs,showkeys]{revtex4}
\usepackage{graphicx,amsmath,amsfonts,amssymb}
\usepackage{times}
\begin{document}
\title{\Large Many-to-one remote information concentration for qudits and multipartite entanglement}
\author{Xin-Wen Wang,$^{1,2,}$\footnote{xwwang@mail.bnu.edu.cn} Shi-Qing Tang,$^{1,2}$ Li-Jun
Xie,$^1$ Deng-Yu Zhang,$^1$ and Le-Man
Kuang$^{2}$\footnote{lmkuang@hunnu.edu.cn}}
\affiliation{$^1$Department of Physics and Electronic Information
Science, Hengyang Normal University, Hengyang 421008, People's Republic of China\\
  $^2$Key Laboratory of Low-Dimensional Quantum Structures and Quantum Control of Ministry of Education,
and Department of Physics, Hunan Normal University, Changsha 410081,
People's Republic of China}

\begin{abstract}
 Telecloning and its reverse process, referred to as
remote information concentration (RIC), have attracted considerable
interest because of their potential applications in
quantum-information processing. We here present a general scheme for
RIC in $d$-level systems (qudits), in which the quantum information
initially distributed in many spatially separated qudits can be
remotely and deterministically concentrated to a single qudit via an
entangled channel without performing any global operations. We show
that the entangled channel of RIC can be different types of
entangled states, including mixed states as well as pure ones. More
interestingly, these mixed states include a bound entangled state
which has a similar form to the generalized Smolin state but has
different characteristics from it. We also show that there exists a
multipartite entangled state which can be used to implement both
telecloning and RIC in the two-level system. Our many-to-one RIC
protocol could be slightly modified to perform some types of
many-to-many RIC tasks.
\end{abstract}



 \maketitle

\section{Introduction}
Quantum mechanics prohibits an unknown quantum state from being
perfectly copied \cite{299N802,92PLA271}. However, an unknown
quantum state can be copied approximately with a certain fidelity
\cite{54PRA1844,5QIC583,7QIC551}, referred to as (approximate)
quantum cloning. Furthermore, when an unknown state comes from a
restricted set of quantum states, it can be faithfully cloned with a
certain probability \cite{80PRL4999,83PRL2849,82PRA062307}, referred
to as probabilistic quantum cloning. Since the seminal work of
Bu\v{z}ek and Hillery \cite{54PRA1844}, quantum cloning has
attracted considerable attention (see
Refs.~\cite{77RMP125,1301.2956} for a review), due to the fact that
it has wide potential applications in quantum-information science as
well as could help us understand quantum mechanics itself more well
(see, e.g.,
\cite{62PRA022301,73PRA032304,11123396,95PRL090504,63PRA042308,82PRA053814,82PRA032325,76PRA042126}).
Although the fidelities of clones relative to the original state are
less than one, the quantum information of the input system is not
degraded but only distributed into a larger quantum system. That is,
the quantum cloning process can be regarded as the distribution of
quantum information from an initial system to final ones. Thus,
quantum cloning combined with remote quantum-information processing
(QIP) may have potential applications in multiparty quantum
communication and distributed quantum computation. This leads to the
advent of the concept of quantum telecloning \cite{59PRA156}, which
is the combination of quantum cloning and quantum teleportation
\cite{70PRL1895}, and functions as simultaneously distributing the
copies of an unknown quantum state to spatially separated sites,
i.e., realizing nonlocal quantum cloning, via a previously shared
multipartite entangled state. Telecloning has been widely studied
and many idiographic schemes have been presented, including
$1\rightarrow N$ telecloning of an arbitrary state or a
phase-covariant state
\cite{59PRA156,61PRA032311,7QIC716,79PRA062315,72PRA032331,67PRA012323,369PLA112,79PRA064306,27CPL100303}.
As the reverse process of telecloning, remote information
concentration (RIC) was first introduced by Murao and Vedral
\cite{86PRL352}. They demonstrated that the quantum information
originally distributed into three spatially separated qubits from a
single qubit can be remotely concentrated back to a single qubit via
a four-qubit unlockable bound entangled state (UBES)
\cite{63PRA032306,80PRL5239,1QIC45} (the four-qubit UBES was first
found by Smolin and is referred to as \emph{Smolin state} or
\emph{Smolin UBES}) without performing any global operations.
Telecloning and RIC processes could be regarded as, respectively,
remote information depositing and withdrawing processes, or remote
information encoding and decoding processes, which is expected to
find useful applications in network-based QIP \cite{86PRL352}. A
scheme for the reverse process of $1\rightarrow 2$ telecloning via a
four-qubit Greenberger-Horne-Zeilinger (GHZ) state \cite{58AJP1131}
has also been proposed \cite{68PRA024303}. Not long before, schemes
for the reverse process of $1 \rightarrow N$ telecloning in
two-level systems have been presented
\cite{73PRA012318,76PRA032311}. Recently, the reverse process of $1
\rightarrow 2$ telecloning in multilevel systems has been studied by
part of our authors \cite{84PRA042310}.

In this paper, we present a general scheme for implementing the
reverse process of $1\rightarrow N$ telecloning of an arbitrary
quantum state in $d$-level systems, which are applicable to
arbitrary $N\geq 2$ and $d\geq 2$ in principle. It will be shown
that the RIC scheme relies on the establishment of special
multiparticle entangled states that function as multiuser
quantum-information channels. Particularly, the quantum channel of
RIC can be different types of entangled states, including mixed
states as well as pure ones; more interestingly, these pure states
include different classes of genuine multipartite entangled states
which are inequivalent under local operations and classical
communication (LOCC), and these mixed states include an UBES which
has a similar form to the generalized Smolin UBES
\cite{73PRA012318,71PRA062317} but has different features from it.
All these entangled states have $d^2$ common commuting stabilizers.
In addition, we show that there exists a multiqubit ($d=2$)
entangled state which can be utilized to implement both telecloning
and RIC. We also discuss the possibility of generalizing our
many-to-one RIC protocol to perform some types of many-to-many RIC
tasks.

\section{Many-to-one RIC in multilevel systems}

\subsection{A brief review of $1\rightarrow N$ universal telecloning}
\noindent
 Before describing our RIC protocol in the next section, we
here briefly summarize the forward process, $1\rightarrow N$
universal telecloning \cite{61PRA032311}. The telecloning scheme
aims at simultaneously distributing the optimal clones of an
arbitrarily unknown qudit state
\begin{equation}
\label{single} |\varphi\rangle_t=\sum_{j=0}^{d-1}x_j|j\rangle_t
\end{equation}
($\sum_{j=0}^{d-1}|x_j|^2=1$) from a distributor (Alice) to $N$
spatially separated receivers (Bob$_1$, Bob$_2$, $\cdots$, Bob$_N$)
with only LOCC. The quantum channel (resource state) can be the
$(2N)$-qudit entangled state
\begin{equation}
\label{telecloning}
   |\Phi\rangle_{t'12\cdots NA_1A_2\cdots A_{N-1}}=\frac{1}{\sqrt{d}}\sum\limits_{j=0}^{d-1}|j\rangle_{t'}|\phi_j\rangle_{12\cdots NA_1A_2\cdots A_{N-1}},
\end{equation}
where
\begin{eqnarray}
\label{phi}
 |\phi_j\rangle_{12\cdots NA_1A_2\cdots A_{N-1}}=&\sum\limits_{n_j=1}^{N}\alpha_{n_j}|\{0,n_0\},\{1,n_1\},\cdots,\{j,n_j\},\cdots,\{d-1,n_{d-1}\}\rangle_{12\cdots
N}\nonumber\otimes\\
&|\{0,n_0\},\{1,n_1\},\cdots,\{j,n_j-1\},\cdots,\{d-1,n_{d-1}\}\rangle_{A_1A_2\cdots
A_{N-1}},
\end{eqnarray}
$\alpha_{n_j}=\sqrt{\frac{n_jd!(N-1)!}{(N+d-1)!}}$
\cite{14EPJB669,64PRA064301}, and
$|\{0,n_0\},\{1,n_1\},\cdots,\{j,n_j\},\cdots,\{d-1,n_{d-1}\}\rangle$
denotes a completely symmetric (normalized) state with $n_j$
particles in the state $|j\rangle$ and $\sum_{j=0}^{d-1}n_j=N$. Here
particle $t'$ is on the sender Alice's hand, particle $s$ is held by
the $s$th recipient Bob$_s$ ($s=1,2,\cdots,N$), and the ancillary
particles $\{A_1,A_2,\cdots,
 A_{N-1}\}$ are arbitrarily distributed among these parties (or even be placed elsewhere). For example, when $N=2$
 the state in Eq.~(\ref{phi}) can be explicitly expressed as
\begin{equation}
  |\phi_j\rangle_{12A}=
  \frac{1}{\sqrt{2(d+1)}}\sum\limits_{r=1}^{d-1}(|j\rangle_1|\overline{j+r}\rangle_2+|\overline{j+r}\rangle_1|j\rangle_2)|\overline{j+r}\rangle_A
  +\sqrt{\frac{2}{d+1}}|j\rangle_1|j\rangle_2|j\rangle_A,
\end{equation}
where $\overline{j+r}=j+r$ modulo $d$. Using the results
\begin{eqnarray}
  &&R^{m,n}|\overline{j+n}\rangle=\omega^{jm}|j\rangle,\nonumber\\
  &&R^{m,n}=\sum_{j=0}^{d-1}\omega^{jm}|j\rangle\langle \overline{j+n}|,
\end{eqnarray}
with $\omega=e^{2\pi i/d}$, it is easy to prove that
\begin{eqnarray}
\label{phitransformation}
 && R^{m,n}_1\otimes
 R^{m,n}_2\cdots  R^{m,n}_N\otimes R^{-m,n}_{A_1}\otimes R^{-m,n}_{A_2}\cdots R^{-m,n}_{A_{N-1}}|\phi_{\overline{j+n}}\rangle_{12\cdots NA_1A_2\cdots A_{N-1}}\nonumber\\
&& =\omega^{jm}|\phi_j\rangle_{12\cdots NA_1A_2\cdots A_{N-1}}.
\end{eqnarray}

The state of the whole system of the $2N+1$ particles
$|\Psi\rangle_{tt'1\cdots NA_1\cdots
A_{N-1}}=|\varphi\rangle_t\otimes|\Phi\rangle_{t'12\cdots
NA_1A_2\cdots A_{N-1}}$ can be expressed as
\begin{eqnarray}
|\Psi\rangle_{tt'1\cdots NA_1\cdots
A_{N-1}}=\frac{1}{d}\sum\limits_{m,n=0}^{d-1}|B^{m,n}\rangle_{tt'}
   \sum\limits_{j=0}^{d-1}\omega^{-jm}x_j|\phi_{\overline{j+n}}\rangle_{12\cdots NA_1A_2\cdots A_{N-1}},
\end{eqnarray}
where $\{|B^{m,n}\rangle: m,n=0,1,\cdots,d-1\}$ are the generalized
Bell-basis states given by
\begin{eqnarray} \label{Bell}
  &&|B^{0,0}\rangle=\frac{1}{\sqrt{d}}\sum\limits_{j=0}^{d-1}|j\rangle|j\rangle,\nonumber\\
  &&|B^{m,n}\rangle=I\otimes U^{m,n}|B^{0,0}\rangle,\nonumber\\
  && U^{m,n}=\sum\limits_{k=0}^{d-1}\omega^{km}|\overline{k+n}\rangle\langle k|.
\end{eqnarray}
The telecloning can now be accomplished by the following simple
procedure: (i) Alice performs a generalized Bell-basis measurement
(GBM) on particles $t$ and $t'$, obtaining one of the $d^2$ outcomes
$\{(m,n): m,n=0,1,\cdots,d-1\}$, and informs all Bobs of the
outcome; (ii) Depending on Alice's outcome $(m,n)$, each Bob
performs a local operation $R^{m,n}$ on his particle. According to
Eq.~(\ref{phitransformation}), if every ancillary particle is also
made a corresponding local operation $R^{-m,n}$, the particles
$\{1,2,\cdots,N\}$ and $\{A_1,A_2,\cdots,A_{N-1}\}$ end in the state
\begin{equation}
\label{clone} |\psi\rangle_{12\cdots NA_1A_2\cdots
A_{N-1}}=\sum\limits_{j=0}^{d-1}x_j|\phi_j\rangle_{12\cdots
NA_1A_2\cdots A_{N-1}}.
\end{equation}
According to Ref.~\cite{84PRA034302}, it can be easily verified that
the collective output state of $N$ clones $\rho^{out}_{N}$ and the
individual output state of one clone $\rho^{out}_{1}$ are the same
as that of Refs.~\cite{14EPJB669,58PRA1827}. Thus, each Bob finally
obtains a clone with the optimal fidelity $F=(2N+d-1)/N(d+1)$. It is
worth pointing out that the local operations on the ancillary
particles are not necessary since the individual output state of a
particle is not related to the local operations on the other
particles.

\subsection{A general scheme for RIC}

In this section, we describe the reverse process of the
aforementioned telecloning, i.e., RIC. After telecloning operations,
the initial single-particle ($t$) quantum information is remotely
distributed into $2N-1$ spatially separated particles
($1,2,\cdots,N,A_1,A_2,\cdots,A_{N-1}$), represented by the
collective quantum state in Eq.~(\ref{clone}). The ownership of
particles $1,2,\cdots,N$ is the same as the preceding section; i.e.,
they are still held by Bob$_1$, Bob$_2$, $\cdots$, Bob$_{N}$,
respectively. Without loss of generality, we assume particles
$A_1,A_2,\cdots,A_{N-1}$ are held by Charlie$_1$, Charlie$_2$,
$\cdots$, Charlie$_{N-1}$, respectively. The RIC is aim to
concentrate the information initially distributed in
$(2N-1)$-particle cloning state of Eq.~(\ref{clone}) back to a
remote particle ($N'$, held by Diana) with only LOCC:
$|\psi\rangle_{12\cdots NA_1A_2\cdots
A_{N-1}}\rightarrow|\varphi\rangle_{N'}$.

In order to show clearly the RIC process and how to construct the
entangled channel, we rewrite the cloning state in Eq.~(\ref{clone})
as (see Appendix A)
\begin{equation}
\label{clone1} |\psi\rangle_{12\cdots NA_1A_2\cdots
A_{N-1}}=\frac{1}{\sqrt{d}}\sum\limits_{m,n=0}^{d-1}\beta_n
|\overline{B_{mn}}\rangle_{12\cdots N-1,A_1A_2\cdots
A_{N-1}}U_{N}^{-m,n}|\varphi\rangle_N,
\end{equation}
where $\sum_{n=0}^{d-1}\beta_n^2=1$ and
\begin{equation}
\label{Bmn}|\overline{B_{mn}}\rangle_{1\cdots N-1,A_1\cdots
A_{N-1}}=\sum\limits_{j_1,\cdots,j_{2N-2}=0}^{d-1}\sqrt{P_{j_1\cdots
j_{2N-2}}}|B^{j_1,j_2}\rangle_{1A_1}\cdots|B^{j_{2N-3},j_{2N-2}}\rangle_{N-1,A_{N-1}}
\end{equation}
with the constraints
\begin{equation}
\sum\limits_{s=1}^{N-1}j_{2s-1}~\mathrm{mod}~d=m,~~~~\sum\limits_{s=1}^{N-1}j_{2s}~\mathrm{mod}~d=n.
\end{equation}
Note that particle $N$ in Eq.~(\ref{clone1}) can be interchanged
with any one of particles $1,2,\cdots,N-1$ because of the
permutability of them.

We first consider employing the following $2N$-particle entangled
pure state as the quantum channel (resource state):
\begin{eqnarray}
\label{channelg} &&|\Psi^g\rangle_{A'_11'A'_22'\cdots A'_N
N'}=\sum\limits_{k_1,\cdots,k_{2N}=0}^{d-1}\sqrt{P_{k_1\cdots
k_{2N}}}|B^{k_1,k_2}\rangle_{A'_11'}\cdots|B^{k_{2N-1},k_{2N}}\rangle_{A'_NN'},\nonumber\\
&&\sum\limits_{s=1}^N
k_{2s-1}~\mathrm{mod}~d=u,~~~~\sum\limits_{s=1}^N
k_{2s}~\mathrm{mod}~d=v,
\end{eqnarray}
where $u$ and $v$ are two arbitrarily given nonnegative integers
that are less than $d$. We assume that particles
$1',2',\cdots,(N-1)',A'_N$ are held by Bob$_1$, Bob$_2$, $\cdots$,
Bob$_{N}$, respectively; particles $A'_1,A'_2,\cdots,A'_{N-1}$ are
held by Charlie$_1$, Charlie$_2$, $\cdots$, Charlie$_{N-1}$,
respectively; particle $N'$ belongs to Diana. A schematic picture of
the RIC protocol is shown in Fig.~1. The procedure is as follows.
(S1) All Bobs and Charlies perform GBMs on their own particles,
respectively. (S2) Each of them tells Diana the measurement outcome
by sending $2\log d$ bits of classical information. (S3) Diana
performs a conditional local operation on particle $N'$.

\begin{figure}
\center
  \includegraphics{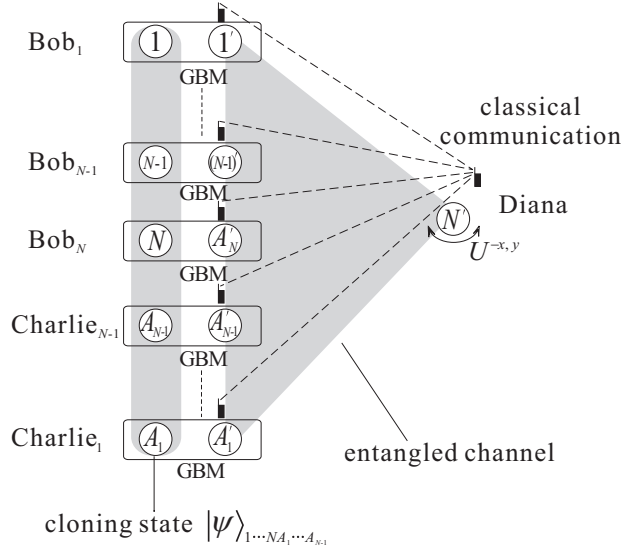}
 \caption{Schematic picture showing the concentration
of information from $N$ Bobs (Bob$_1$, Bob$_2$, $\cdots$, Bob$_N$)
and $N-1$ Charlies (Charlie$_1$, Charlie$_2$, $\cdots$,
Charlie$_{N-1}$) to the remote receiver, Diana, using an entangled
channel. The gray rectangle denotes the cloning state of $2N-1$
qudits, the gray triangle denotes the entangled channel, the blank
rectangles denote the generalized Bell-state measurements, and the
dashed lines mean the classical communications.}
\end{figure}

In (S1), the GBMs of all Bobs and Charlies are independent, and thus
the sequence can be arbitrary. For clarity, we here assume that
Bob$_N$ performs the GBM after the others. Based on the identity
\begin{equation}
\label{swapping}
|B^{m,n}\rangle_{XY}|B^{m',n'}\rangle_{X'Y'}=\frac{1}{d}\sum\limits_{m'',n''=0}^{d-1}
 \omega^{m''n''}|B^{\overline{m+m''},\overline{n'+n''}}\rangle_{XY'}|B^{\overline{m'-m''},\overline{n-n''}}\rangle_{X'Y}
\end{equation}
with $\overline{m'-m''}=m'-m''+d~\mathrm{modulo}~ d$, we can obtain
the relationship of the measurement outcomes of these parties.
Without loss of generality, we particularly assume the measurement
outcomes of Bob$_s$ and Charlie$_s$ ($s=1,2,\cdots,N-1$) are
$(\overline{j_{2s-1}+l_{2s-1}},\overline{k_{2s}+l_{2s}})$ and
$(\overline{k_{2s-1}-l_{2s-1}},\overline{j_{2s}-l_{2s}})$,
respectively. Then Diana can obtain the results
\begin{eqnarray}
&&\sum\limits_{s=1}^{N-1}(j_{2s-1}+k_{2s-1})~\mathrm{mod}~d=\left(m+\sum\limits_{s=1}^{N-1}k_{2s-1}\right)~\mathrm{mod}~d=u',\nonumber\\
&&\sum\limits_{s=1}^{N-1}(j_{2s}+k_{2s})~\mathrm{mod}~d=\left(n+\sum\limits_{s=1}^{N-1}k_{2s}\right)~\mathrm{mod}~d=v'.
\end{eqnarray}
As a consequence, $N$, $A'_N$, and $N'$ are projected in the state
\begin{eqnarray}
\label{teleportation}
&&U_{N}^{-m,n}|\varphi\rangle_N|B^{k_{2N-1},k_{2N}}\rangle_{A'_NN'}\nonumber\\
&&=\frac{1}{d}\sum\limits_{x,y=0}^{d-1}\omega^{n(m-k_{2N-1})+(k_{2N}-n)x}
|B^{\overline{x+k_{2N-1}-m},\overline{y+k_{2N}-n}}\rangle_{NA'_N}U_{N'}^{-x,y}|\varphi\rangle_{N'}.
\end{eqnarray}
Next Bob$_N$ performs a GBM on particles $N$ and $A'_N$, which can
be regarded as being equivalent to Bob$_N$ and Diana together
performing the teleportation protocol with a local error-correction
operation on particle $N'$. Assume that the outcome is
$(u''=\overline{x+k_{2N-1}-m},v''=\overline{y+k_{2N}-n})$ and
particle $N'$ is projected in the state
$U_{N'}^{-x,y}|\varphi\rangle_{N'}$. After receiving all the
measurement outcomes sending from the other participants, Diana can
deduce the result $(x=\overline{u''+u'-u}, y=\overline{v''+v'-v})$.
Then, Diana performs the local operation
$(U_{N'}^{-x,y})^+=R_{N'}^{x,y}$ and obtains the state
$|\varphi\rangle_{N'}$. As a consequence, the information initially
distributed in $2N-1$ spatially separated particles is now remotely
concentrated in a single particle.

Equation (\ref{channelg}) contains a broad family of entangled pure
states. We now consider some special cases. Assuming $k_2\equiv
k_4\equiv\cdots\equiv k_{2N}\equiv0$, $P_{k_1\cdots k_{2N}}\equiv
1/d^{N-1}$, and $u=0$, the state in Eq.~(\ref{channelg}) reduces to
(see Appendix B)
\begin{equation}
\label{channels1}
 |\Psi^{s_1}\rangle_{A'_11'A'_22'\cdots A'_NN'}
 =\frac{1}{\sqrt{d}}\sum\limits_{j=0}^{d-1}|j\rangle_{A'_1}|j\rangle_{1'}|j\rangle_{A'_2}|j\rangle_{2'}\cdots|j\rangle_{A'_N}|j\rangle_{N'},
\end{equation}
i.e., a generalized GHZ state \cite{63PRA022104}. Assuming
$\sum\limits_{s=1}^{N-1} k_{2s-1}~\mathrm{mod}~d=x$, $k_{2N-1}=d-x$,
$\sum\limits_{s=1}^{N-1}k_{2s}~\mathrm{mod}~d=y$, $k_{2N}=d-y$, and
$\sqrt{P_{k_1\cdots k_{2N}}}=\beta_y\sqrt{P_{k_1\cdots
k_{2N-2}}}/\sqrt{d}$ [the definition of $\beta_y$ is the same as
Eq.~(\ref{clone1})], the entangled channel in Eq.~(\ref{channelg})
reduces to
\begin{equation}
\label{channels2}
 |\Psi^{s_2}\rangle_{A'_11'A'_22'\cdots A'_NN'}=\frac{1}{\sqrt{d}}\sum\limits_{x,y=0}^{d-1}\beta_y
|\overline{B_{xy}}\rangle_{A'_11'A'_22'\cdots
A'_{N-1}(N-1)'}|B^{\overline{-x},\overline{-y}}\rangle_{A'_NN'}.
\end{equation}
For the case $d=2$ (qubit), we proved that the state of
Eq.~(\ref{channels2}) is the same as that of Eq.~(\ref{telecloning})
(see Appendix C). This indicates that the multiqubit entangled state
in Eq.~(\ref{telecloning}) can be competent for implementing both
telecloning and RIC, two inverse processes. In other words, the
aforementioned telecloning and RIC for $d=2$ can be achieved by
using the same entangled channel. However, such a result is not
applicable to $d>2$ (qudit). This is an interesting difference
between qudit-RIC and qubit-RIC. According to Ref.~\cite{59PRA156},
the states of Eqs.~(\ref{channels1}) and (\ref{channels2}) with
$d=2$ are not equivalent to each other, i.e., cannot be transformed
into each other by LOCC. It can be verified that the states of
Eqs.~(\ref{channels1}) and (\ref{channels2}) with $d>2$ are also
LOCC inequivalent. This implies that Eq.~(\ref{channelg}) contains
at least two inequivalent classes of genuine $2N$-partite entangled
pure states. In other words, different classes of genuine $2N$-qudit
entangled pure states can implement a same multiparty QIP task,
$(2N-1)\rightarrow 1$ RIC. Such a phenomenon is counterintuitive,
since a given QIP task can be achieved by only typical structure of
entangled states and different types of entangled states are usually
competent for implementing different QIP tasks. It has been shown
\cite{74PRA062320,8QIP431} that quantum teleportation can be
deterministically implemented by using both multiqubit W and GHZ
states, two inequivalent genuine multiqubit entangled states
\cite{62PRA062314}. However, teleportation is a two-party
communication, and the W and GHZ states in fact play the same role
as the bipartite entangled state, i.e., only the bipartite
entanglement of them is exploited. In contrast, RIC is a multiparty
communication (each party holds one particle of the entangled
channel), and the states of Eqs.~(\ref{channels1}) and
(\ref{channels2}) play a role of multipartite entanglement.

We now show that the quantum channel of our RIC can also be a broad
family of entangled mixed states. Let $P_{k_1\cdots
k_{2N}}=\delta_{k_1,c_1}\cdots \delta_{k_{2N},c_{2N}}$, where
$c_1,\cdots,c_{2N}$ are arbitrarily chosen nonnegative integers that
are less than $d$. Then the entangled channel in
Eq.~(\ref{channelg}) reduces to a product state of $N$ generalized
Bell states,
\begin{equation}
\label{channels3}
  |\Psi^{s_3}\rangle_{A'_11'A'_22'\cdots A'_NN'}=|B^{c_1,c_2}\rangle_{A'_11'}|B^{c_3,c_4}\rangle_{A'_22'}\cdots|B^{c_{2N-1},c_{2N}}\rangle_{A'_NN'}.
\end{equation}
Because the constants $c_1,\cdots,c_{2N}$ are arbitrary, we deduce
that the quantum channel of our RIC can also be the following form
of entangled mixed states:
\begin{equation}
\label{channelrho}
 \rho_{A'_11'\cdots
A'_NN'}=\sum\limits_{k_1,\cdots,k_{2N}=0}^{d-1}C_{k_1\cdots
k_{2N}}|B^{k_1k_2}\rangle_{A'_11'}\langle B^{k_1k_2}| \otimes\cdots
\otimes|B^{k_{2N-1}k_{2N}}\rangle_{A'_NN'}\langle
B^{k_{2N-1}k_{2N}}|,
\end{equation}
where $\sum C_{k_1\cdots k_{2N}}=1$.
 This can be easily proved by resorting to a purified state of
$\rho_{A'_11'\cdots A'_NN'}$,
\begin{equation}
 |\Psi^{\rho}\rangle_{A'_11'\cdots A'_NN'}=\sum\limits_{k_1,\cdots,k_{2N}=0}^{d-1}\sqrt{C_{k_1\cdots
k_{2N}}}|B^{k_1,k_2}\rangle_{A'_11'}\cdots|B^{k_{2N-1},k_{2N}}\rangle_{A'_NN'}|\mathcal{A}_{k_1\cdots
k_{2N}}\rangle_A,
\end{equation}
where $\{|\mathcal{A}_{k_1\cdots k_{2N}}\rangle_A\}$ are orthogonal
normalized states of an ancillary system $A$. Particularly, by
carrying out the same procedure as before [see (S1)-(S3)], the
information of $|\psi\rangle_{1\cdots NA_1\cdots A_{N-1}}$ can also
be concentrated in particle $N'$ via the entangled channel
$|\Psi^{\rho}\rangle_{A'_11'\cdots A'_NN'}$. In the whole process,
the ancillary system $A$ is not touched, and thus can be traced out
at any time. This finishes the proof that the mixed state
$\rho_{A'_11'\cdots A'_NN'}$ can be used to implement our RIC.

If we set $u=v=0$ and $C_{k_1\cdots k_{2N}}=1/d^{2(N-1)}$,
Eq.~(\ref{channelrho}) reduces to
\begin{equation}
\label{channelrho'}
 \rho'_{A'_11'\cdots A'_NN'}=\frac{1}{d^{2(N-1)}}\sum\limits_{k_1,\cdots,k_{2N}=0}^{d-1}|B^{k_1k_2}\rangle_{A'_11'}\langle
B^{k_1k_2}| \otimes\cdots
\otimes|B^{k_{2N-1}k_{2N}}\rangle_{A'_NN'}\langle
B^{k_{2N-1}k_{2N}}|.
\end{equation}
For $d=2$, $\rho'_{A'_11'\cdots A'_NN'}$ is exactly the generalized
Smolin state \cite{73PRA012318,71PRA062317}, a $2N$-qubit UBES. The
generalized Smolin UBES is fully symmetric; i.e., it is unchanged
under permutation of any two qubits. This leads to the generalized
Smolin UBES being separable with respect to any $2:2$ partition of
$\{A'_1,A'_2,\cdots,A'_N,1',2',\cdots,N'\}$. For $d>2$,
$\rho'_{A'_11'\cdots A'_NN'}$ also describes an UBES (see Appendix
D); and when $N=2$ it recovers the results of
Ref.~\cite{84PRA042310}. However, $\rho'_{A'_11'\cdots A'_NN'}$ with
$d>2$ is an asymmetric but not symmetric UBES, because
$\{A'_1,A'_2,\cdots,A'_N,1',2',\cdots,N'\}$ are not completely
permutable, i.e., particularly, any one of
$G_1=\{A'_1,A'_2,\cdots,A'_N\}$ and any one of
$G_2=\{1',2',\cdots,N'\}$ are not permutable, as shown in Appendix
D. The asymmetry results in the fact that $\rho'_{A'_11'\cdots
A'_NN'}$ cannot be superactivated for $d>2$, which presents a
striking contrast to the generalized Smolin UBES being
superactivable \cite{76PRA022322}. These results indicate that there
exists an analog to the generalized Smolin UBES in multilevel
systems; however, it has some different characteristics. Note that
the asymmetric $2N$-qudit UBES didn't appear in previous literature,
and thus is a ``new'' asymmetric UBES.

As shown above, different types of entangled states, including both
pure and mixed states, can be exploited as the quantum channel of
many-to-one RIC. The pure states can be multiple-Bell states and
LOCC inequivalent genuine $2N$-partite entangled states. The mixed
states could even be bound entangled states. However, it can be
verified that all these states have several common properties as
follows. (a) All of them are stabilized by the Abelian group
$S=\{S^{mn}=\otimes_{s=1}^{N}U^{-m,n}_{A'_s}\otimes U^{m,n}_{s'}:
m,n=0,1,\cdots, d-1\}$; that is, for any $m$ and $n$,
$\mathrm{tr}(S^{mn}|\Psi^g\rangle_{A'_11'\cdots
A'_NN'}\langle\Psi^g|)=\mathrm{tr}(S^{mn}\rho_{A'_11'\cdots
A'_NN'})=1$. (b) They can be expanded by the generalized Bell states
with the same constraints [see the second row of
Eq.~(\ref{channelg})]. (c) The amount of entanglement across the
$\{A'_1,A'_2,\cdots,A'_N,1',2',\cdots,(N-1)'\}:\{N'\}$ cut is
$\log_2d$ ebit, which ensures that the success probability of
remotely concentrating one-qudit information is one.

The key points for the physical or experimental realization of the
RIC task above are as follows: (i) preparation of the entangled
channel, i.e., the generalized Bell states or GHZ states, or the
UBES of Eq.~(\ref{channelrho'}); (ii) realization of telecloning (or
cloning) of an arbitrary quantum state; (iii) implementation of the
GBM. All these building blocks are achievable in quantum optics as
discussed in Ref.~\cite{84PRA042310}.

\section{Discussion and conclusion}

A more general RIC protocol should be a many-to-many protocol.
However, it will be much more complicated and cannot be obtained by
directly generalizing the many-to-one protocol shown above. As a
matter of fact, there are two types of many-to-many RIC protocols.
One involves more than one receiver. The other aims at concentrating
multi-qudit information to a remote site. For the former case, we
here consider the reverse process of the ``many-to-many'' quantum
information distribution presented in Ref.~\cite{67PRA012323}. In
the ``many-to-many'' information distribution protocol of
Ref.~\cite{67PRA012323}, the information of an entangled state
\begin{equation}
|\varphi'\rangle=\sum\limits_{j=0}^{d-1}x_j|j\rangle_{t_1}|j\rangle_{t_2}\cdots
|j\rangle_{t_L}
\end{equation}
shared by $L$ spatially separated distributors is transmitted by
telecloning procedure to $M$ receivers ($M>L$) situated at different
locations. Naturally, the reverse process of it is to remotely
concentrate the information distributed in $M$ particles back to $L$
spatially separated particles. Let $M=2N-1$, this task can be
implemented by slightly modifying the aforementioned many-to-one RIC
protocol, with the high-dimensional Bell state
$|B^{k_{2N-1},k_{2N}}\rangle_{A'_NN'}$ in Eqs.~(\ref{channelg}),
(\ref{teleportation}), and (\ref{channelrho}) being replaced by the
high-dimensional GHZ state
\begin{eqnarray}
&& |G^{k_{2N-1},k_{2N}}\rangle=I\otimes U^{k_{2N-1},k_{2N}}\otimes \underbrace{U^{0,k_{2N}}\otimes\cdots\otimes U^{0,k_{2N}}}\limits_{L-1}|G^{0,0}\rangle_{A'_NN'_1N'_2\cdots N'_L},\nonumber\\
&& |G^{0,0}\rangle_{A'_NN'_1N'_2\cdots N'_L}=\frac{1}{\sqrt{d}}\sum\limits_{j=0}^{d-1}|j\rangle_{A'_N}|j\rangle_{N'_1}|j\rangle_{N'_2}\cdots |j\rangle_{N'_L},\nonumber\\
&& U^{k_{2N-1},k_{2N}}=\sum\limits_{j=0}^{d-1}\omega^{jk_{2N-1}}|\overline{j+k_{2N}}\rangle\langle j|,
\end{eqnarray}
and $|\varphi\rangle_{N'}$ and $U^{-x,y}_{N'}$ in
Eq.~(\ref{teleportation}) replaced by
$|\varphi'\rangle_{N'_1N'_2\cdots N'_L}$ and $U^{-x,y}_{N'_1}\otimes
U^{0,y}_{N'_2}\otimes\cdots\otimes U^{0,y}_{N'_L}$, respectively.

As a direct extension of the aforementioned many-to-one RIC, the
second type of many-to-many RIC, which aims at concentrating
multi-qudit information to a remote site, should be the reverse
process of $L\rightarrow N$ ($N>L$) optimal universal telecloning
\cite{87PRA022302,47JMO247}. However, it is not clear whether the
output state of $L\rightarrow N$ optimal universal cloning has the
form similar to that in Eq.~(\ref{clone1}). Thus we cannot construct
the entangled channel by the idea similar to that used in our
many-to-one RIC protocol and choose suitable operations. We here
discuss alternatively a simple scenario, i.e., the reverse process
of the following many-to-many quantum information distribution.
Suppose that Alice had distributed the information of $L$ identical
but unknown $d$-level quantum states $|\varphi\rangle^{\otimes L}$
into a $(2N-L)$-qudit state
\begin{equation}
  \label{cloneL} |\psi\rangle_{\{2N-L\}}=\frac{1}{\sqrt{d}}\sum\limits_{m,n=0}^{d-1}\beta_n
|\overline{B_{mn}}\rangle_{\{2N-2L\}}\left(U^{-m,n}\right)^{\otimes
L}|\varphi\rangle^{\otimes L}
\end{equation}
shared by $(2N-L)$ spatially separated clients. Note that this state
is not necessarily to be the output state of the so-called
$L\rightarrow N$ ($N>L$) optimal universal telecloning
\cite{87PRA022302,47JMO247}. The reverse process is to remotely
concentrate the distributed information in $2N-L$ spatially
separated particles back to $L$ particles held by a receiver. It is
easy to verify that such a RIC task can be accomplished by the same
procedure as the aforementioned many-to-one RIC via the quantum
channel $|B^{0,0}\rangle^{\otimes N}$ shared among the $(2N-L)$
senders (each one holds one particle of a Bell state) and a receiver
(holds $L$ particles of $L$ Bell states).

In conclusion, we have studied the many-to-one RIC, i.e., the
reverse process of $1\rightarrow N$ universal telecloning, in
$d$-level systems, which are applied to arbitrary $N \geq 2$ and $d
\geq 2$ in principle. We have shown that the quantum channel of RIC
can be different types of entangled states, including mixed states
as well as pure ones, in contrast to telecloning which requires a
certain type of entangled channel. Such a difference may be due to
the fact that RIC can be considered to be a disentangling operation,
whereas telecloning can be considered to be an entangling operation.
Although these entangled states are LOCC inequivalent, they have a
common feature, i.e., have $d^2$ common commuting stabilizers. We
have also revealed concomitantly some interesting entanglement
phenomena as follows. (a) Similar to qubit-RIC, qudit-RIC can also
be implemented by an UBES. Though such a multilevel UBES has a
similar form to the generalized Smolin UBES, it has some different
features; particularly, the former one has asymmetry and the latter
one has symmetry. (b) Telecloning and RIC for qubits can be achieved
by using the same entangled channel, but there is no such feature
for qudits. Our many-to-one RIC protocol can be slightly modified to
implement some many-to-many RIC tasks. These protocols are
experimentally achievable in the field of quantum optics.

Subsequent to submitting this manuscript, Zhang \emph{et al.}
independently proposed a many-to-one RIC protocol with the
generalized Bell states acting as the entangled channel
\cite{87PRA022302}. This paper has shown that many-to-one RIC can be
realized by different channels including both pure and mixed
entangled states (even bound entangled states). As a matter of fact,
the entangled channel used in Ref.~\cite{87PRA022302} is the same as
that in Eq.~(\ref{channels3}) with $c_1=c_2=\cdots=c_{2N}=0$, i.e.,
a special case of the general channel in Eq.~(\ref{channelg}), of
the present paper.

\begin{acknowledgments}
This work was supported by the National Natural Science Foundation
of China (Grant Nos. 11004050 and 11075050), the Program for
Changjiang Scholars and Innovative Research Team in University
(Grant No. IRT0964), the Key Project of Chinese Ministry of
Education (Grant No. 211119), the China Postdoctoral Science
Foundation funded project (Grant No. 2012M511729), the Hunan
Provincial Natural Science Foundation (Grant Nos. 11JJ7001 and
11JJ6063), and the construct program of the key discipline in Hunan
province.
\end{acknowledgments}

\appendix
\section{}
 In this appendix, we demonstrate that the cloning state of
Eq.~(\ref{clone}) can be rewritten as form of Eq.~(\ref{clone1}). To
satisfy Eq.~(\ref{phitransformation}), $|\phi_j\rangle$ can be
rewritten as
\begin{equation}
\label{Aphi} |\phi_j\rangle_{12\cdots NA_1A_2\cdots
A_{N-1}}=\sum\limits_{n=0}^{d-1}\beta_n|\lambda_{j_n}\rangle_{12\cdots
N-1,A_1A_2\cdots A_{N-1}}|\overline{j+n}\rangle_{N},
\end{equation}
where
\begin{eqnarray}
\label{lambdatransformation}
 &&R^{k,l}_1\otimes
 R^{k,l}_2\cdots  R^{k,l}_{N-1}\otimes R^{-k,l}_{A_1}\otimes R^{-k,l}_{A_2}\cdots R^{-k,l}_{A_{N-1}}|\lambda_{j_n}\rangle_{12\cdots N-1,A_1A_2\cdots A_{N-1}}\nonumber\\
&& =\omega^{-nk}|\lambda_{\overline{(j-l)}_n}\rangle_{12\cdots
N-1,A_1A_2\cdots A_{N-1}}.
\end{eqnarray}
Now let
\begin{equation}
\label{Bmn1}
|\overline{B_{mn}}\rangle=\frac{1}{\sqrt{d}}\sum\limits_{j=0}^{d-1}\omega^{jm}|\lambda_{j_n}\rangle.
\end{equation}
It can be verified that
\begin{eqnarray}
\label{Btransformation}
 &&R^{k,l}_1\otimes
 R^{k,l}_2\cdots  R^{k,l}_{N-1}\otimes R^{-k,l}_{A_1}\otimes R^{-k,l}_{A_2}\cdots R^{-k,l}_{A_{N-1}}|\overline{B_{mn}}\rangle_{12\cdots N-1,A_1A_2\cdots A_{N-1}}\nonumber\\
&& =\omega^{lm-nk}|\overline{B_{mn}}\rangle_{12\cdots
N-1,A_1A_2\cdots A_{N-1}}.
\end{eqnarray}
We notice that
\begin{equation}
R^{k,l}\otimes
R^{-k,l}|B^{x,y}\rangle=\omega^{lx-yk}|B^{x,y}\rangle.
\end{equation}
Therefore, $|\overline{B_{mn}}\rangle_{12\cdots N-1,A_1A_2\cdots
A_{N-1}}$ can also be expressed as the form of Eq.~(\ref{Bmn}). From
Eq.~(\ref{Bmn1}), we obtain
\begin{equation}
\label{lambda}
|\lambda_{j_n}\rangle=\frac{1}{\sqrt{d}}\sum\limits_{m=0}^{d-1}\omega^{-jm}|\overline{B_{mn}}\rangle.
\end{equation}
Then Eq.~(\ref{clone1}) can be obtained by substituting
Eqs.~(\ref{Aphi}) and (\ref{lambda}) into Eq.~(\ref{clone}).

\section{}
 If $k_2\equiv k_4\equiv\cdots \equiv k_{2N}\equiv0$,
$P_{k_1\cdots k_{2N}}\equiv1/d^{N-1}$, and $u=0$, the state in
Eq.~(\ref{channelg}) can be expressed as
\begin{eqnarray}
\label{APsis1}
 |\Psi^{s_1}\rangle_{A'_11'A'_22'\cdots A'_NN'}&=&
\frac{1}{\sqrt{d^{N-1}}}\sum\limits_{k_3,k_5,\cdots,k_{2N-1}=0}^{d-1}|B^{-k_3-k_5\cdots
-k_{2N-1},0}\rangle_{A'_1,1'}\nonumber\\
&&\otimes|B^{k_{3},0}\rangle_{A'_2,2'}\otimes\cdots\otimes|B^{k_{2N-1},0}\rangle_{A'_N,N'}\nonumber\\
&=&\frac{1}{\sqrt{d^{2N-1}}}\sum\limits_{k_3,k_5,\cdots,k_{2N-1}=0}^{d-1}\sum\limits_{j_1,j_3,\cdots,j_{2N-1}=0}^{d-1}\omega^{j_1(-k_3-k_5\cdots
-k_{2N-1})}|j_1\rangle_{A'_1}|j_1\rangle_{1'}\nonumber\\
&& \otimes
\omega^{j_3k_3}|j_3\rangle_{A'_2}|j_3\rangle_{2'}\otimes\cdots
\otimes
\omega^{j_{2N-1}k_{2N-1}}|j_{2N-1}\rangle_{A'_N}|j_{2N-1}\rangle_{N'}\nonumber\\
&=&\frac{1}{\sqrt{d^{2N-1}}}\sum\limits_{j_1,j_3,\cdots,j_{2N-1}=0}^{d-1}|j_1\rangle_{A'_1}|j_1\rangle_{1'}|j_3\rangle_{A'_2}|j_3\rangle_{2'}\cdots|j_{2N-1}\rangle_{A'_N}|j_{2N-1}\rangle_{N'}\nonumber\\
&& \times
\sum\limits_{k_3=0}^{d-1}\omega^{(j_1-j_3)k_3}\sum\limits_{k_5=0}^{d-1}\omega^{(j_1-j_5)k_5}\cdots
\sum\limits_{k_{2N-1}=0}^{d-1}\omega^{(j_1-j_{2N-1})k_{2N-1}}\nonumber\\
&=&
\frac{1}{\sqrt{d}}\sum\limits_{j_1=0}^{d-1}|j_1\rangle_{A'_1}|j_1\rangle_{1'}|j_1\rangle_{A'_2}|j_1\rangle_{2'}\cdots|j_1\rangle_{A'_N}|j_1\rangle_{N'}.
\end{eqnarray}
Here we have used the identity
$\sum_{k=0}^{d-1}\omega^{jk}=d\delta_{j,0}$, where
$\delta_{j=0,0}=1$ and  $\delta_{j\neq 0,0}=0$. Obviously, the state
of Eq.~(\ref{APsis1}) is the same as that of Eq.~(\ref{channels1}),
i.e., a normal generalized GHZ state.

\section{}
This appendix shows the equivalence of the state in
Eq.~(\ref{telecloning}) to the state in Eq.~(\ref{channels2}) for
$d=2$. By substituting Eqs.~(\ref{Aphi}) and (\ref{lambda}) into
Eq.~(\ref{telecloning}), the telecloning state
$|\Phi\rangle_{t'12\cdots NA_1A_2\cdots A_{N-1}}$ reads
\begin{eqnarray}
\label{APhi} |\Phi\rangle_{t'12\cdots NA_1A_2\cdots
A_{N-1}}&=&\frac{1}{d}\sum\limits_{j=0}^{d-1}|j\rangle_{t'}\sum\limits_{y=0}^{d-1}\beta_{y}\sum\limits_{x=0}^{d-1}\omega^{-jx}|\overline{B_{xy}}\rangle_{12\cdots
N-1,A_1A_2\cdots
A_{N-1}}|j+y\rangle_{N}\nonumber\\
&=&\frac{1}{d\sqrt{d}}\sum\limits_{x,y,j=0}^{d-1}\beta_{y}\omega^{-jx}|\overline{B_{xy}}\rangle_{12\cdots
N-1,A_1A_2\cdots
A_{N-1}}\sum\limits_{l=0}^{d-1}\omega^{-lj}|B^{l,y}\rangle_{t'N}\nonumber\\
&=&\frac{1}{d\sqrt{d}}\sum\limits_{x,y,j,l=0}^{d-1}\beta_{y}\omega^{-j(x+l)}|\overline{B_{xy}}\rangle_{12\cdots
N-1,A_1A_2\cdots
A_{N-1}}\sum\limits_{l=0}^{d-1}|B^{l,y}\rangle_{t'N}\nonumber\\
&=&\frac{1}{\sqrt{d}}\sum\limits_{x,y=0}^{d-1}\beta_{y}|\overline{B_{xy}}\rangle_{12\cdots
N-1,A_1A_2\cdots A_{N-1}}|B^{\overline{-x},y}\rangle_{t'N}.
\end{eqnarray}
Here we have used the identity
\begin{equation}
\label{equality}
  |j\rangle_{t'}|k\rangle_{N}=\frac{1}{\sqrt{d}}\sum\limits_{l=0}^{d-1}\omega^{-jl}|B^{l,\overline{k-j}}\rangle_{t'N}~~~~(0\leq j,k\leq
  d-1),
\end{equation}
which can be obtained from Eq.~(\ref{Bell}). For $d=2$,
Eq.~(\ref{APhi}) reduces to
\begin{eqnarray}
\label{APhi1} |\Phi\rangle_{t'12\cdots NA_1A_2\cdots
A_{N-1}}&=&\frac{1}{\sqrt{d}}\sum\limits_{x,y=0}^{1}\beta_{y}|\overline{B_{xy}}\rangle_{12\cdots
N-1,A_1A_2\cdots
A_{N-1}}|B^{\overline{-x},y}\rangle_{t'N}\nonumber\\
&=&\frac{1}{\sqrt{d}}\sum\limits_{x,y=0}^{1}\beta_{y}|\overline{B_{xy}}\rangle_{12\cdots
N-1,A_1A_2\cdots
A_{N-1}}|B^{\overline{-x},\overline{-y}}\rangle_{t'N},
\end{eqnarray}
which is obviously the same as the state of Eq.~(\ref{channels2})
with $d=2$.

\section{}
We here prove that the state $\rho'_{A'_11'A'_22'\cdots
A'_NN'}$ in Eq.~(\ref{channelrho'}) is an asymmetric UBES for any
$d> 2$, by using some results of Ref.~\cite{75PRA052332}. We define
an Abelian subgroup of the generalized Pauli group
\cite{75PRA052332},
\begin{equation}
\label{Abelian}
S=\{S^{mn}=\bigotimes\limits_{s=1}^{N}U^{-m,n}_{A'_s}\otimes
U^{m,n}_{s'}: m,n=0,1,\cdots, d-1\},
\end{equation}
which is composed of $d^2$ commuting operators. A state
$|\psi\rangle$ is said to be stabilized by $S$, if
$S^{mn}|\psi\rangle=1,~\forall~m,n=0,1,\cdots,d-1$. All the states
stabilized by $S$ constitute a subspace, denoted by $H_S$, of the
Hilbert space of $n$ qudits. Define $T_s=\{A'_s,s'\}$
($s=1,2,\cdots,N$) and $S^{mn}_{T_s}=U^{-m,n}_{A'_s}\otimes
U^{m,n}_{s'}$. It can be verified that any two operators
$S^{mn}_{T_s}$ and $S^{m'n'}_{T_s}$ are commutable,
$\forall~s=1,2,\cdots,N$. Then the two operators $S^{mn},S^{m'n'}\in
S$ are said to commute locally with respect to the partition
$\{T_1,T_2,\cdots,T_N\}$ of
$\{A'_1,A'_2,\cdots,A'_N,1',2',\cdots,N'\}$, and $S$ is said to be
separable with respect to this partition \cite{75PRA052332}.

It can be verified that
\begin{equation}
S^{mn}_{T_s}|B^{x_s,y_s}\rangle_{A'_ss'}=\omega^{y_sm-x_sn}|B^{x_s,y_s}\rangle_{A'_ss'},
\end{equation}
$\forall~s=1,2,\cdots,N$; i.e.,
$\{|B^{x_s,y_s}\rangle_{A'_ss'}:x_s,y_s=0,1,\cdots,d-1\}$ are the
simultaneous eigenstates of $S^{mn}_{T_s}$ corresponding to the
eigenvalues $\{\omega^{y_sm-x_sn}:x_s,y_s=0,1,\cdots,d-1\}$ for each
$m,n=0,1,\cdots,d-1$. Then it is obvious that the $2N$-qudit states
$\{\otimes_{s=1}^{N}|B^{x_s,y_s}\rangle_{A'_s,s'}:x_s,y_s=0,1,\cdots,d-1\}$
are the simultaneous eigenstates of $S^{mn}$ with the eigenvalues
$\{\omega^{\sum_{s=1}^{N}y_sm-\sum_{s=1}^{N}x_sn}:x_s,y_s=0,1,\cdots,d-1\}$
for each $m,n=0,1,\cdots,d-1$. In particular, each term of the state
$\rho'_{A'_11'A'_22'\cdots A'_NN'}$ in Eq.~(\ref{channelrho'}) is
the simultaneous eigenstate of $S^{mn}$ with eigenvalue 1 for each
$m,n=0,1,\cdots,d-1$. These eigenstates also form an orthonormal
basis of the stabilized space $H_S$. According to \emph{Lemma 1} of
Ref.~\cite{75PRA052332}, the state $\rho'_{A'_11'A'_22'\cdots
A'_NN'}$ in Eq.~(\ref{channelrho'}) is the maximally mixed state
over $H_S$.

As have been shown that $S$ is separable with respect to the
partition $\{T_1,T_2,\cdots,T_N\}$. It can also be verified that for
any $X\neq Y\in\{A'_1,A'_2,\cdots,A'_N,1',2',\cdots,N'\}$, there
exists at least one partition $\{g_1,g_2,\cdots,g_f\}$ with $X\in
g_1$, $Y\in g_2$ such that $S$ is separable with respect to this
partition. These results satisfy the condition 1 in \emph{Theorem 1}
of Ref.~\cite{75PRA052332}, which indicates that
$\rho'_{A'_11'A'_22'\cdots A'_NN'}$ is a bound entangled state. The
unlockability or activability of $\rho'_{A'_11'A'_22'\cdots A'_NN'}$
is obvious. For example, it can be unlocked as follows: let qudits
$A'_s$ and $s'$ ($s=2,\cdots,N$) join together and perform a GBM on
them; then depending on the measurement outcome qudits $A'_1$ and
$1'$ is projected in a generalized Bell state, i.e., pure
entanglement is distilled out between qudits $A'_1$ and $1'$. In
fact $S_{T_s}=\{S^{mn}_{T_s}:m,n=0,1,\cdots,d-1\}$ is obviously
inseparable, $\forall~s=1,2,\cdots,N$, which satisfies the condition
2 in \emph{Theorem 1} of Ref.~\cite{75PRA052332}. Thus
$\rho'_{A'_11'A'_22'\cdots A'_NN'}$ is an UBES.

We now classify the $2N$ qudits of the state
$\rho'_{A'_11'A'_22'\cdots A'_NN'}$ into two groups
$G_1=\{A'_1,A'_2,\cdots,A'_N\}$ and $G_2=\{1',2',\cdots,N'\}$. It is
obvious that $S$ acts symmetrically on the $N$ qudits of each group,
which indicates that the state remains invariant when exchanging any
two qudits inside the same group. However, when we exchange two
qudits that belong to two different groups, the state will change.
Therefore, the UBES $\rho'_{A'_11'A'_22'\cdots A'_NN'}$ is
asymmetric when $d>2$.

\end{document}